\documentclass[preprint,showpacs]{revtex4}
\usepackage{graphics}

\newcommand{\nl}{\nonumber \\}

\newcommand{\be}{\begin{equation}}
\newcommand{\ee}{\end{equation}}
\newcommand{\bea}{\begin{eqnarray}}
\newcommand{\eea}{\end{eqnarray}}

\newcommand{\Eq}[1]{Eq.\,(\ref{#1})}

\begin{document}

\title{Theoretical investigation of the quantum noise 
in ghost imaging
}

\author{Jing Cheng, Shensheng Han}

\affiliation{ Key Laboratory for Quantum Optics 
and Center for Cold Atom Physics, 
Shanghai Institute of Optics and Fine Mechanics, 
Chinese Academy of Sciences, Shanghai, 201800, P.R.China
}

\begin{abstract}
Ghost imaging is a method to nonlocally image an 
object by transmitting pairs of entangled photons through the object 
and a reference optical system respectively. We present a 
theoretical analysis of the quantum noise in this imaging 
technique. The dependence of the noise on the properties of the 
apertures in the imaging system are discussed
and demonstrated with a numerical example. For a given source,  
the resolution and the signal-to-noise ratio cannot be improved 
at the same time . 
\end{abstract}

\maketitle

In recent years, there has been an increasing interest in the 
topic of ghost imaging \cite{jetp,initexp,
pra2000,josab2002,retrodic,shih04,macro,corr}. 
Ghost imaging is a kind 
of correlated imaging and relies on 
the quantum entanglement of the photon pairs created 
by spontaneous parametric down conversion (SPDC).
The photons of a pair are spatially separated. One of them 
propagates through a known (reference) optical imaging system, 
while the other travels through an unknown (test) optical 
imaging system in which an object is placed. By measuring 
the coincidence rate of these photon pairs at the reference and 
test detectors, one can obtain the image of the object as a 
function of the transverse position of the reference photon.
The theory of ghost imaging was firstly studied by Klyshko 
\cite{jetp}. Then the experiments were demonstrated in mid-1990s 
\cite{initexp}. A systematic theory was given by the Boston
group in \cite{pra2000,josab2002}. 
Gatti {\it et al.} generalized the theory 
to deal with the case in which the number of entangled photons 
are large \cite{macro}.

Resolution and noise are two of the most important factors 
to characterize an imaging system. Resolution of a ghost imaging 
system has been investigated in some papers. 
In references \cite{josab2002} and 
\cite{retrodic}, the authors have analyzed the dependence of 
imaging resolution on some physical parameters.
However, the noise in ghost imaging attracts little attention. 
In this paper, we give a
theoretical investigation of the quantum noise in ghost imaging.
Based on the general theory given in \cite{pra2000,josab2002},
we derive a formula to calculate the variance of the 
coincidence rate. Especially, our main interest is to study the 
dependence of the quantum noise 
on the impulse respond function of both 
reference and test imaging systems. For simplicity, the possible 
fluctuations from the detectors are not considered.
We find that, when the imaging system 
has a large cutoff frequency (corresponding to good resolution), 
the variance of the coincidence rate may be very large and leads 
to bad signal-to-noise ratio (SNR).

\begin{figure}[ht]
{\scalebox{.9}{\includegraphics*{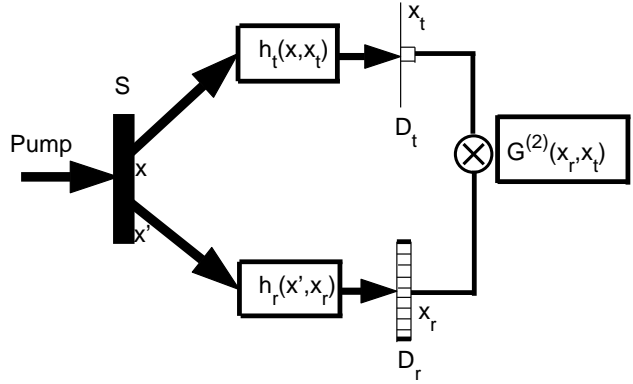}}}
\caption{A setup of entangled ghost imaging.
A pump field incidents on a nonlinear crystal $S$.
The source $S$ emits pairs of entangled photons. One of the photons 
transmits through the test system $h_t(x,x_t)$ which contains an 
unknown object, and the other photon transits the reference system 
with a known $h_r(x',x_r)$. Two detectors $D_t$ and $D_r$ record 
the intensity distribution. The coincidence rate 
$G^{(2)}(x_r,x_t)$ is measured to give a image of the object.
}
\label{setup}
\end{figure}

Let's consider the setup of a ghost imaging system given in  
Fig (\ref{setup}). The source $S$ produces pairs of entangled 
photons. These photon pairs are transmitted through a reference 
optical system and a test optical system which contains the 
object to be imaged. These two optical systems are characterized 
by their impulse response function $h_r(x,x_r)$ and $h_t(x,x_t)$ 
respectively. Two detectors $D_1$ and $D_2$ record the intensity 
distribution of the test and reference photons. 
The coincidence rate of photon pairs at these 
two detectors ($G^{(2)}(x_r,x_t)$) is proportional to 
the fourth-order correlation function \cite{pra2000,josab2002}
\be
G^{(2)}(x_r,x_t)=
\left| {\int\!\!\!\int {dxdx'} \varphi (x,x')
       h_t (x_t ,x)h_r (x_r,x')} \right|^2   ,
\label{G2}
\ee
where $\varphi(x,x')$ represents the wave function of
entangled photons.

For simplicity, we define an operator
\be
\hat S (x_t,x_r)=\hat E_t^-(x_t) \hat E_r^-(x_r) 
      \hat E_t^+(x_t) \hat E_r^+(x_r)     ,
\label{Soper}
\ee
where $\hat E_t^-(x_t)$, $\hat E_r^-(x_r)$,
$\hat E_t^+(x_t)$, $\hat E_r^+(x_r)$ are operators for the 
negative and positive frequency portions of the test and 
reference photons at the positions $x_t$ and $x_r$. 
As shown in Ref \cite{pra2000}, the probability of coincidence of
photons at the positions $x_t$ and $x_r$ is the expected 
value of operator $\hat S$
\be
G^{(2)}(x_r,x_t)=
\left\langle \Psi  \right| \hat S(x_t,x_r)
          \left| \Psi  \right\rangle ,
\label{G2oper}
\ee

The photon pairs can be described by the pure 
two-photon state as
\be
\left| \Psi  \right\rangle  = \int\!\!\!\int {dxdx'} \varphi 
(x,x') \hat a_t^\dag(x) \hat a_r^\dag(x') 
\left| {0,0} \right\rangle ,
\label{Psioper}
\ee
where $\left| {0,0} \right\rangle$ is the vacuum state, 
$a$ ane $a^\dag$ are creation operators for the test and reference
photons at position x and x'.
The relations between $\hat E_t^+(x_t)$, $\hat E_r^+(x_r)$ and 
the operators at the source are \cite{pra2000}
\bea
\hat E_t^+(x_t) &=& \int\!dx_t\, h_t(x_t,x) \hat a_t(x) , 
\label{Et} \\
\hat E_r^+(x_r) &=& \int\!dx_t\, h_r(x_r,x') \hat a_r(x') .
\label{Er} 
\eea
\Eq{G2} is derived by directly substituting 
Eqs. (\ref{Psioper},\ref{Et},\ref{Er}) into \Eq{G2oper}.

To study the quantum noise in ghost imaging, we need to calculate the 
variance of the operator $\hat S$. The quantum fluctuation 
of the coincidence rate $\Delta G^{(2)}$ is obtained from 
\be
\Delta G^{(2)}(x_r,x_t)
=  \sqrt{  \left\langle \Psi  \right|
            [\hat S(x_t,x_r)]^2 \left| \Psi  \right\rangle 
          -[G^{(2)}(x_r,x_t)]^2 
        }   .
\label{DG2oper}
\ee
By direct calculations, we can obtain a mathematic formula for
$\left\langle \Psi  \right|
[\hat S(x_t,x_r)]^2 \left| \Psi  \right\rangle $,
\bea
\left\langle \Psi  \right|
[\hat S(x_t,x_r)]^2 \left| \Psi  \right\rangle
&=&
  G^{(2)}(x_r,x_t) \times 
  \int {dx} \left|{ h_t (x_t ,x) }\right|^2  \nl 
&&  \times \int {dx'} \left|{ h_r (x_r ,x') }\right|^2 .
\label{SS}
\eea
\Eq{DG2oper} and \Eq{SS} are the main equations in this paper.

Now, we give some discussions on the quantum noise in ghost imaging.
Firstly, when the coincidence rate ($G^{(2)}$) is increasing, 
the noise ($\Delta G^{(2)}$) is also increasing, but the 
signal-to-noise ratio (SNR) will be enhanced,
\bea
{\rm SNR} &=& \frac{G^{(2)}(x_r,x_t)}{\Delta G^{(2)}(x_r,x_t)} \nl
&=& 
\frac{1}{ \sqrt{
   \frac{\int {dx} \left|{ h_t (x_t ,x) }\right|^2
    \int {dx'} \left|{ h_r (x_r ,x') }\right|^2}{G^{(2)}(x_r,x_t)}
  - 1 } } .
\label{snr}
\eea

Further, the noise also depends on the two impulse response 
functions. Since $\left\langle \Psi  \right|
[\hat S(x_t,x_r)]^2 \left| \Psi  \right\rangle $ is proportional
to the integral of $\left|{ h_t}\right|^2$ and 
$\left|{ h_r}\right|^2$, one needs to carefully design the optical 
imaging systems to control the noise.

An optical imaging system may contain different optical 
elements, such as lens and apertures. Generally, all apertures 
can be projected through an effective exit aperture characterized 
by its pupil function $p(x_a)$. 
As discussed in Fourier optics \cite{fo}, for a coherent 
imaging system, the relation between the image ($u_i(x_i)$)
and the object ($u_o(x_o)$) is
\be
u(x_i)=\int {dx} h(x_i-x_o)u_o(Mx_o) , 
\label{cohimg}
\ee 
where $M$ is the magnification of the system. The impulse 
response function is given by
\be
h(x)={\bf F}[p(\eta x)],
\ee
where ${\bf F}[f(x)]$ means the Fourier transformation of 
function $f(x)$ and $\eta$ is a constant depends on the geometric
parameters. In many cases, 
these impulse response functions are shift-invariant. Then
the integral $\int {dx_o} \left|{ h (x_i ,x_o) }\right|^2$ 
is determined by the cutoff frequency of the optical transfer 
function. Since the cutoff frequency is proportional to the 
size of the aperture, this integral is proportional to the areas of 
the exit aperture. Thus, the noise will be small if the exit 
aperture is small. However, small aperture means small cutoff
frequency, so the image resolution will be decreased. On the 
other hand, using a large aperture can increase the image 
resolution, but the noise is also increased. So good resolution 
and small noise can not be realized at the same time.

In practical experiments, thousands of photon pairs are used to 
get an image. Suppose there are $N$ entangled photon 
pairs which are generated independently, then
the fluctuation of the averaged coincidence rate will be 
$\sqrt{N}$ times smaller and
the SNR will be enhanced by a factor $\sqrt{N}$.

We give a numerical example to see the dependence on the 
apertures. The test imaging system
consists of an object characterized by a 
transmission function $t(x)$, a lens, and a detector ($D_t$). 
The lens is located at a focal distance $f$ from the object and 
from $D_t$. If the size of the lens is much larger than the 
object, $h_t$ has the form
\be
h_t(x_t,x)=-\frac{i}{\lambda f}t(x)
     \exp \left(-\frac{2\pi i}{\lambda f}x_t  x \right)  ,
\label{ht}
\ee
where $\lambda$ is the wavelength.
In the reference imaging system, a lens is placed at a distance 
$2f$ both from the source and the detector $D_r$. 
Then $h_r$ has the form
\be
h_r(x_r,x')=\frac{1}{4\lambda^2 f^2}
     P\left(\frac{x_r+x'}{2\lambda f} \right)
     \exp \left(\frac{i \pi }{2\lambda f} (x_r^2+x^2) \right)  ,
\label{hr}
\ee
where $P(u)$ is the Fourier transformation of the pupil function
of the lens $p(x)$. Such a kind of ghost imaging system can 
image the object in the reference detector \cite{josab2002,corr}. 

\begin{figure}[ht]
{\scalebox{.9}{\includegraphics*{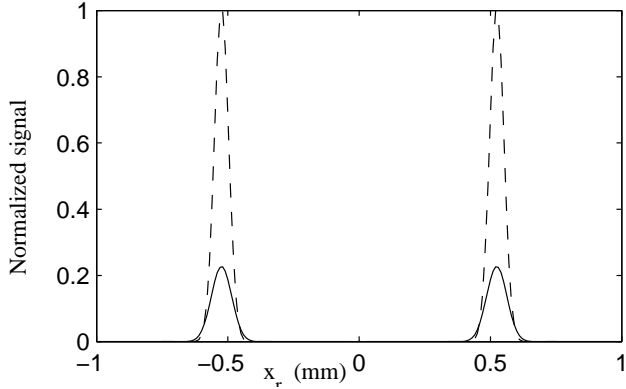}}}
\caption{ Normalized conditional coincidence rate 
$G^{(2)}(x_r,0)$ (dashed line) and 
Normalized quantum fluctuation of the conditional 
coincidence rate $\Delta G^{(2)}(x_r,0)$ (solid line)
as the function of $x_r$. A large aperture is used ($D=10{\rm mm}$).
}
\label{S0}
\end{figure}

A double slit is used in our calculation as the object, with 
the width of the two slits $w=0.05{\rm mm}$ and the distance between 
them $d=1{\rm mm}$. The size of the lens is $D$. $f=100{\rm mm}$,
$\lambda=650{\rm nm}$ and $N=10000$ are 
used in our calculation. The precise 
formula of $\varphi(x,x')$ may be very complicated, as given in
\cite{pra2000},
\be
\varphi(x,x') \propto \int {dy} E_p(y)\zeta (x-y,x'-y) ,
\ee
where $E_p(x)$ is the pump field and $\zeta(x,x')$ is 
a phase-matching function depends on the crystal parameters.
Here we use a simplified formula
\be
\varphi(x,x')=C\exp\left(-\frac{x^2+x'^2}{a^2} \right)
       \exp\left(-\frac{(x-x')^2}{b^2} \right)  ,
\label{phi}
\ee
where parameter $a=2{\rm mm}$ decides the size of the source, 
$b=0.05{\rm mm}$ determines the degree of the entanglement, 
$C$ is a normalized constant. 

First, for a large aperture $D=10{\rm mm}$, 
in Fig (\ref{S0}), the dashed line shows the conditional coincidence 
rate ($G^{(2)}(x_r,0)$) and the solid line is
the quantum fluctuation of the conditional coincidence rate
($\Delta G^{(2)}(x_r,0) $).
These two curves are normalized with 
the maximum value of $G^{(2)}(x_r,0)$.
The resolution is not bad and the two slits are very clear.
Also the noise is not very large, the SNR is about $4$. So 
the image quality is good in ghost imaging for given parameters.

\begin{figure}[ht]
{\scalebox{.9}{\includegraphics*{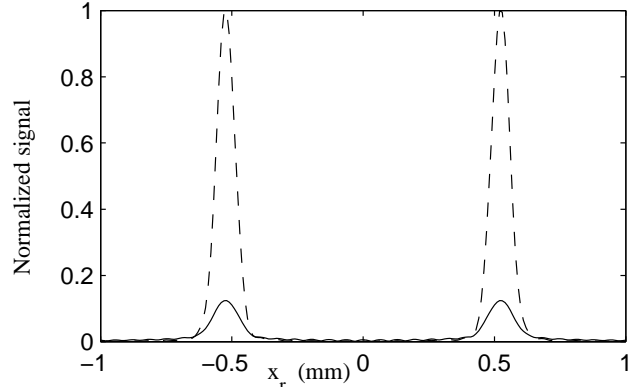}}}
\caption{ The same as in Fig (\ref{S0}), but $D=2{\rm mm}$.
}
\label{Sc}
\end{figure}

As we have discussed before, the aperture of the imaging system can 
affect the image quality. In Fig (\ref{Sc}),
we use a small aperture lens with $D=2{\rm mm}$. Small lens make 
the resolution degrade. But, as shown in Fig (\ref{Sc}), 
the amplitude of the noise is decreased about two times 
compared with Fig (\ref{S0}). 
To obtain a good image, we need to balance the 
requirements on resolution and noise in a ghost imaging system.

In conclusion, the quantum noise in entangled 
ghost imaging has been investigated 
theoretically in this paper. We have 
presented the mathematical formula to calculate the quantum 
fluctuation of the coincidence rate suitable for various imaging 
configurations. 
Apertures in the imaging system affect the imaging quality 
significantly. Using small apertures will decrease the noise but 
also degrade the resolution. It is impossible to improve both 
the resolution and SNR at the same time by designing the 
imaging system only.

The first author acknowledges the supports from the China 
Postdoctoral Science Foundation and the K.C. Wong Education 
Foundation, Hong Kong.

\end{document}